**Title:** Econophysics: historical perspectives
**Contributors:** Gilles Daniel and Didier Sornette
**Affiliation:** ETH Zurich, Chair of Entrepreneurial Risks, Department of Management, Technology and Economics, Zurich

**Abstract:** *Econophysics* embodies the recent upsurge of interest by physicists into financial economics, driven by the availability of large amount of data, job shortage in physics and the possibility of applying many-body techniques developed in statistical and theoretical physics to the understanding of the self-organizing economy. This brief historical survey emphasizes that *Econophysics* has many historical precursors, and is in fact rooted in a continuous cross-fertilization between economics and physics that has been active in the last centuries.

**Main text**
The term *Econophysics* was introduced circa 1994, endorsed in 1999 by the publication of its founding book, Mantegna-Stanley's "An Introduction to Econophysics" (1999). The word "econophysics" suggests that there is a physical approach to economics, perhaps even that economics can be rooted in physics, paralleling the quests of biophysics or geophysics.

Indeed, all along its developments, from classical to neo-classical economics and till the present time, economists have been inspired by the conceptual and mathematical developments of the physical sciences and by their remarkable successes in describing and predicting natural phenomena. Reciprocally, physics has been enriched several times by developments first observed in economics. Well before the christening of econophysics as the incarnation of the multidisciplinary study of complex large-scale financial and economic systems, a multiple of small and large collisions have punctuated the development of these two fields. Let us now mention a few that illustrate the remarkable commonalities and inter-fertilization.

In his "Inquiry into the Nature and Causes of the Wealth of Nations" (1776), Adam Smith found inspiration in the Philosophiae Naturalis Principia Mathematica (1687) of Isaac Newton, specifically based on the (novel at the time) notion of causative forces.

The recognition of the importance of feedbacks to fathom the sheer complexity of economic systems has been at the root of economic thinking for a long time. Towards the end of the 19$^{th}$ century, the microeconomists Francis Edgeworth and Alfred Marshall drew on some of the ideas of physicists to develop the notion that the economy achieves an equilibrium state like that described for gases by Clerk Maxwell and Ludwig Boltzmann. The general equilibrium theory now at the core of much of economic thinking is nothing but a formalization of the idea that "everything in the economy affects everything else" (Krugman, 1996), reminiscent of mean-field theory or self-consistent effective medium methods in physics, but emphasizing and transcending these ideas much beyond their initial sense in physics.

While developing the field of microeconomics in his "Cours d'Economie Politique" (1897), the economist and philosopher Vilfredo Pareto was the first to describe, for the distribution of incomes, the eponym power-laws that would later become the center of attention of Physicists and other scientists observing this remarkable and universal statistical signature in the distribution of event sizes (earthquakes, avalanches, landslides, storms, forest fires, solar flares, commercial sales, war sizes, and so on) punctuating so many natural and social systems [Mandelbrot, 1982; Bak, 1996; Newman, 2005; Sornette, 2006].

While attempting to model the erratic motion of bonds and stock options in the Paris *Bourse* in 1900, mathematician Louis Bachelier developed the mathematical theory of diffusion (and the first elements of financial option pricing) and solved the parabolic diffusion equation five years before Albert Einstein (1905) established the theory of Brownian motion based on the same diffusion equation (also underpinning the theory of random walks). The ensuing modern theory of random walks now constitutes one of the fundamental pillars of theoretical physics and economics and finance models.

In the early 1960s, mathematician Benoit Mandelbrot (1963) pioneered the use in Financial Economics of heavy-tailed distributions (Lévy stable laws) as opposed to the traditional Gaussian (Normal) law. A cohort of economists, notably at the University of Chicago (Merton Miller, Eugene Fama, Richard Roll), at MIT (Paul Samuelson) and at Carnegie Mellon University (Thomas Sargent) initially followed his steps. In his PhD thesis, Eugene Fama confirmed that the frequency distribution of the changes in the logarithms of prices was "leptokurtic", i.e., with a high peak and fat tails. However, other notable economists (Paul Cootner and Clive Granger) strongly opposed Mandelbrot's proposal, based on the argument that "the statistical theory that exists for the normal case is nonexistent for the other members of the class of Lévy laws." The coup-de-grace was the mounting empirical evidence that the distributions of returns were becoming closer to the Gaussian law at time scales larger than one month, at odds with the self-similarity hypothesis associated with the Lévy laws (Campbell et al., 1997; MacKenzie, 2006). Much of the efforts in the econophysics literature of the late 1990s and early 2000s revisited and refined this hypothesis, confirming on one hand the existence of the variance (which rules out the class of Lévy distributions proposed by Mandelbrot), but also suggesting a power law tail with an exponent close to 3 (Mantegna and Stanley, 1995; Gopikrishnan et al., 1999) -- several other groups have discussed alternatives, such as exponential (Silva et al. (2004) or stretched exponential distributions (Laherrere and Sornette, 1999;

Malevergne et al., 2005; Malevergne and Sornette, 2006). Financial engineers actually care about these apparent technicalities because the tail structure controls the Value-at-Risk and other measures of large losses, and physicists care because the tail may constrain the underlying mechanism(s). For instance, Gabaix et al. (2003) attribute the large movements in stock market activity to the interplay between the power-law distribution of the sizes of large financial institutions and the optimal trading of such large institutions. In this domain, econophysics focuses on models that can reproduce and explain the main stylized facts of financial time series: non-Gaussian fat tail distribution of returns, long-range auto-correlation of volatility and absence of correlation of returns, multifractal property of the absolute value of returns, and so on.

In the late 1960s, Benoit Mandelbrot left financial economics but, inspired by this first episode, went on to explore other uncharted territories to show how non-differentiable geometries (that he coined "fractal"), previously developed by mathematicians from the 1870s to the 1940s, could provide new ways to deal with the real complexity of the world (Mandelbrot, 1982). He later returned to finance in the late 1990s in the midst of the econophysics' enthusiasm to model the multifractal properties associated with the long-memory properties observed in financial asset returns (Mandelbrot et al., 1997; Mandelbrot, 1997; Bacry et al., 2001; Muzy et al., 2001; Sornette et al., 2003).

The modern econophysicists are implicitly and sometimes explicitly driven by the hope that the concept of "universality" holds in economics and finance. The value of this strategy remains to be validated (Sornette et al., 2007), as most econophysicists have not yet digested the subtleties of economic thinking and failed to marry their ideas and techniques with mainstream economics. The following is a partial list of a few notable exceptions: precursory physics approach to social systems (Galam and Moscovici, 1991), agent-based models, induction, evolutionary models (Farmer, 2002; Arthur, 2005; Cont and Bouchaud, 2000; Lux and Marchesi,1999), option theory for incomplete markets (Bouchaud and Sornette,1994; Bouchaud and Potters, 2003), interest rate curves (Bouchaud et al., 1999; Santa-Clara and Sornette, 2001), minority games (Challet et al., 2005), theory of Zipf law and its economic consequences (Gabaix, 1999, 2005; Malevergne and Sornette, 2007), theory of large price fluctuations (Gabaix et al., 2003), theory of bubbles and crashes (Johansen et al., 1999; Lux and Sornette, 2002; Sornette,2003), random matrix theory applied to covariance of returns (Laloux et al., 1999; Plerou et al.,1999; Pafka and Kondor, 2002), methods and models of dependence between financial assets (Malevergne and Sornette, 2003).

At present, the most exciting progresses seem to be unraveling at the boundary between economics and the biological, cognitive and behavioral sciences. While it is difficult to argue for a physics-based foundation of economics and finance, physics has still a role to play as a unifying framework full of concepts and tools to deal with the complex. The specific training of physicists explains the impressive number of recruitments in investment and financial institutions, where their data-driven approach coupled with a pragmatic sense of theorizing has made physicists a most valuable commodity on Wall Street.

**Acknowledgments:** We would like to thank Y. Malevergne for many discussions and a long-term enjoyable and fruitful collaboration.